\documentstyle[12pt,prl,aps,preprint,tighten,floats]{revtex}
\begin{document}
\draft


\title{String-Loop Corrections Versus Non-Extremality}
\author {Mikhail Z. Iofa\thanks{e-mail: iofa@theory.npi.msu.su}\\
Nuclear Physics Institute \\
Moscow State University\\
Moscow 119899, Russia\\
and\\
Leopoldo A. Pando Zayas\thanks{e-mail:
pandol@uwstout.edu}\\
Physics Department\\
University of Wisconsin-Stout\\
Menomonie, WI 54751, USA}

\def\f0{f_{0}}
\def\pp{\partial_{z}}
\def\ppm{\partial_{\bar{z}}}
\def\pd{\partial_{\mu}}
\def\pa{\partial}
\def\pu{\partial^{\mu}}
\def\P{\Phi}
\def\p{\phi}
\def\L{\Lambda}
\def\S{\Sigma}
\def\l{\lambda}
\def\s{\sigma}
\def\te{\theta}
\def\g{\gamma}
\def\d{\delta}
\def\D{\Delta}
\def\a{\alpha}
\def\b{\beta}
\def\vp{\varphi}
\def\hP{\hat P}
\maketitle

\begin{abstract}
We discuss a magnetic black-hole solution to the equations of motion of the
string-loop-corrected effective action. At the string-tree level, this
solution is the extremal magnetic black hole described by the
"chiral null model." In the extremal case, the string-loop correction
is constant, and this fact is used to  analytically solve
 the loop-corrected equations of motion.
In distinction to the tree-level solution,
the resulting loop-corrected solution has the horizon at a finite distance
from the origin; its location is a function of the loop correction.
The loop-corrected configuration is compared with a string-tree-level
non-extremal magnetic black hole solution
which also has the horizon at a finite distance from the
origin.
We find that for an appropriate choice of free parameters of solutions,
the loop-corrected magnetic black hole can be approximated by  a tree-level
non-extremal solution.
We compare the thermodynamic properties of the loop-corrected and
non-extremal solutions.
\end{abstract}

\pagebreak
\section{Introduction}
At present, string theory is considered the best candidate for
 a fundamental theory that would provide a consistent quantum theory of
gravity unified with the other interactions \cite{gsw,polchinski}.
In particular, string theory provides a powerful approach to the physics of
black holes (review and further refs. in \cite{yo,peet}). In this setting a
fundamental problem is that of
understanding how the intrinsically stringy effects modify  Einstein
gravity. In this paper we discuss two of these effects: presence of
scalar fields such as the dilaton and the moduli fields and higher-genus
contributions modifying  the tree-level effective action.
We focus on higher-genus corrections, because
string theory, being a theory formulated on a world sheet,
intrinsically contains
string-loop corrections from higher topologies of the world sheet (these
vanish only for higher supersymmetries $N\geq 4$), while ${\a}'$ corrections
can vanish  in  certain constructions based on conformal field theories and
for a large class of backgrounds.

We consider
the one-string-loop (torus topology) corrections to a special class of
backgrounds: magnetic $4D$ black holes provided by  the "chiral null
models" embedded in (compactified) heterotic string theory \cite{bh1,bh2}.
 String-loop effects modify the gauge couplings in the effective action.
In the general case, these corrections are rather complicated functions of
the moduli (fields that describe compact dimensions) \cite{ka,uni,kou}.
 However, for the
extremal magnetic black hole solution, the loop correction $\D$ is constant.
Assuming that the loop correction $\D$ is numerically small, we
obtain an analytic solution to the loop-corrected equations of motion.
 The dilaton of the string-tree-level solution increases at small distances.
Thus, at small distances, the effective gauge coupling  $e^{-\phi}
+\D$ is sensitive to string-loop correction. This produces an important
physical effect: the horizon of the loop-corrected magnetic black hole is
shifted from the point $r=0$ to a finite distance determined by the loop
correction.

The loop-corrected solution is compared with a non-extremal string-tree-level
magnetic black hole solution. Since  for small loop
correction  $\D$ the effective string coupling
$e^{-\phi}$ at large distances is much larger
than the loop correction, adjusting free parameters of the non-extremal
solution, the large-$r$ asymptotic of the loop-corrected
metric can be made close to that of the non-extremal solution. Requiring
that the locations of the horizons of both solutions are of the same order,
it is possible to fix the free parameters of both solutions and to
approximate string-loop-corrected black hole solution by a tree-level
non-extremal configuration.

In section 2 we review the structure of the 4D magnetic black hole
in heterotic string theory provided by the chiral null model.

In section 3 we solve the loop-corrected equations of motion. Starting from
the extremal magnetic black-hole solution at the string-tree level, we
obtain an analytic solution to the loop-corrected equations of
motion.

In section 4 we find the position of the horizon of the loop-corrected magnetic
black hole as a function of the loop correction $\D$ and the free
parameters of solution.
 We show that by adjusting the free parameters of both solutions
the loop-corrected solution can be approximated by a
non-extremal string-tree-level magnetic solution.

In section 5 we discuss some thermodynamic properties of the loop-corrected
magnetic black hole. We calculate the Hawking temperature and geometric
entropy of the black hole and compare them with the corresponding quantities
of the non-extremal solution.  Finally, we comment on possible microscopic
derivation of the  entropy.
\section{Dyonic 4D black hole in toroidally compactified heterotic string
theory}
We begin with a brief review of the chiral null models \cite{bh1,bh2}.
The chiral null model
is a nonlinear $2D$ $\s $-model interpreted as a string world-sheet
Lagrangian with nontrivial backgrounds:
\begin{equation}
   L =  F(x)  \pa u \left [\bar{\pa} v +
   K(u,x)\bar{\pa}  u  +   2{\cal A}_i(u,x)  \bar{\pa}  x^i
\right]
 +   (G_{ij} + B_{ij})(x) \pa x^i \bar{\pa} x^j    +    {\cal R}
\Phi (x)\  .
 \label{E1}
\end{equation}
An important special case of the chiral null models (\ref{E1}) are the chiral
null models with  curved transverse part of the form
\begin{eqnarray}
 L = F(x)  \partial u \bigg(\bar{\partial} v +  K(x) \bar{\partial} u\bigg)  +
 f(x)k(x)  \big[ \partial x^4 + a_s (x) \partial x^s\big] \big[
\bar{\pa} x^4  + a_s (x) \bar{\pa} x^s\big]\nonumber \\
    + f(x)  k^{-1} (x) \partial x^s \bar{\partial} x^s
 +   b_s (x) (\partial x^4 \bar{\partial} x^s - \bar{\partial} x^4 \partial
x^s)
  +   {\cal R}   \Phi(x),
 \label{E2}
\end{eqnarray}
$$
{\cal R} \equiv {\textstyle {1\over 4}} \alpha'
\sqrt{ g^{(2)}}  R^{(2)}.
$$
Here $x^s=(x^1,x^2,x^3)$ and $v=2t$ are  non-compact space-time coordinates,
$u=y_2$ and  $x^4=y_1$ are compact toroidal coordinates.
 Written in "$4D$ form", the chiral null model is
$$
L =   (G'_{\mu\nu} + B'_{\mu\nu}) (x) \partial x^\mu \bar{\partial} x^\nu
+
 G_{mn}(x) [ \partial y^m   + A^{(1)m}_{\mu}(x)  \partial x^\mu]
  [ \bar{\partial} y^n   + A^{(1)n}_{\nu}(x)  \bar{\partial} x^\nu]
$$
\begin{equation}
+ \ A^{(2)}_{n\mu}(x)  (\partial y^n \bar{\partial} x^\mu - \bar{\partial}
y^n
\partial x^\mu)
+  {\cal R} \phi(x)  .
\label{E3}
\end{equation}
The $4D$ string-frame background is related to the fields in the Lagrangian
(\ref{E2}) as \cite{sen}
\begin{equation}
G'_{\mu\nu} = G_{\mu\nu} - G_{mn} A^{(1)m}_{\mu} A^{(1)n}_{\nu}, \ \ \
B'_{\mu\nu} = B_{\mu\nu},
\label{E4}
\end{equation}
\begin{equation}
A^{(1)n}_{\mu} \equiv A^{n}_{\mu} =  G^{nm} G_{m\mu} \ , \ \ \
A^{(2)}_{n\mu } =  B_{n\mu}, \ \ \
\phi = \Phi  -{1\over 4}\ln \Delta, \ \ \  \Delta\equiv \det G_{mn},
\label{E5}
\end{equation}
and their explicit form can be read off from (\ref{E2}):
$$
A^{1}_{\mu} = (0,\ a_s), \ \ \ A^{2}_{\mu} = (K^{-1},\ 0),
\ \ \ B_{1 \mu} = (0,\ b_s), \ \ \ B_{2 \mu} = (F,\ 0).
$$
Following \cite{howe} we consider a special class of chiral null models
where  the  functions $ a_s $ and  $ b_s $  satisfy the equations
$$
2\pa_{[p}b_{q]}=-\epsilon_{pqs}\pa^s f, \qquad
2\pa_{[p}a_{q]}=-\epsilon_{pqs}\pa^s k^{-1}.
$$
Here $f$ and $k^{-1}$ are three-dimensional harmonic functions
$${\pa}_s {\pa}^s f=0, \qquad {\pa}_s {\pa}^s k^{-1}=0
$$
and the dilaton $\phi$ is
$$ \phi =\frac{1}{2}\ln f.$$
The four-dimensional Einstein-frame metric is
\begin{equation}
g_{\mu\nu} = e^{-2\phi} G'_{\mu\nu}   .
\label{E6}
\end{equation}

 From the sigma-model perspective,
the $4D$ backgrounds of the above functional form
 are solutions to the equations ensuring conformal invariance of the
theory.
 These backgrounds have also the space-time interpretation
 as black holes that can be obtained by
solving the equations of motion of the $4D$ low energy effective action which
(in the Einstein frame) is
\begin{eqnarray}
\label{E7} S&=&\int d^4
x\sqrt{-g}\left(R-\frac{1}{2}(\pd\p)^2-(\pd\s)^2-(\pd\g)^2 \right. \nonumber
\\ &-&\frac{1}{4}e^{-\p+2\g}F_{(1)}^2-\frac{1}{4}e^{-\p+2\s}F_{(2)}^2
\nonumber \\ &-& \left.  \frac{1}{4}e^{-\p-2\g}F_{(3)}^2
-\frac{1}{4}e^{-\p-2\s}F_{(4)}^2\right).
\end{eqnarray}

Here $g_{\mu \nu}$ is the (Einstein frame) metric, $\s $ and $\g $ are
the moduli related to the radii of the $T^2$.
The action (\ref{E7}) is a particular case of the general 4D effective action
which is obtained by dimensional reduction of 10D supergravity action
\cite{sen} by compactifying on the product of the tori and by truncating to
the relevant set of fields. We shall consider the
following solution to the equations of motion \cite{bh2,cvyo}:
\begin{eqnarray}
\label{E8}
ds^2&=&-{\Lambda}(r)dt^2 +{\Lambda}^{-1}(dr^2+r^2d\Omega_2^2),\nonumber \\
{\Lambda}^2(r)&=&FK^{-1}kf^{-1},\nonumber \\
2\p&=&\ln FK^{-1}fk^{-1}, \nonumber \\
e^{2\s}&=&FK, \nonumber \\
e^{2\g}&=&fk,
\end{eqnarray}
where $F^{-1}, K, f$ and $k^{-1}$ are harmonic functions.
$F_{(1)}$ and $F_{(3)}$ correspond to electric fields, $F_{(2)}$ and
$F_{(4)}$ are magnetic field strengths.

Following the general discussion \cite{ka,uni,kou},
the one-loop-corrected $4D$ action has the
following form:
\begin{eqnarray}
\label{E10}
S&=&\int d^4 x\sqrt{-g}\left(R-\frac{1}{2}(\pd\p)^2-(\pd\s)^2-(\pd\g)^2
\right. \nonumber \\
&-&
\frac{1}{4}e^{2\g}(e^{-\p}+\D)F_{(1)}^2-\frac{1}{4}e^{2\s}(e^{-\p}+\D)
F_{(2)}^2 \nonumber \\
&-&
\left.\frac{1}{4}e^{-2\g}(e^{-\p}+\D)F_{(3)}^2-\frac{1}{4}
e^{-2\s}(e^{-\p}+\D)F_{(4)}^2\right)
\end{eqnarray}
Here the loop correction $\D$ is a function of $G_{11}=fk$ and
$G_{22}=FK$ and in the general case has  a rather complicated form.

Let us consider purely magnetic $\left(F^{-1} = K = 1 \right )$
extremal black holes  $(f^{-1} = k)$.
In this case in (\ref{E8}) $\s = \g =0$ and backgrounds of the
chiral null model are expressed by a single function (we consider
one-center solution) $ f_{0}$:
\begin{eqnarray}
\label{E9}
f_{0}(r)=1+\frac{P}{r}, \nonumber \\
\Lambda =f=k^{-1}= \f0, \nonumber \\
a_\varphi = b_\varphi =P(1-\cos\vartheta),\nonumber \\
\p =\ln \f0 \qquad \g =0,
\end{eqnarray}
where $a_\varphi$ and $b_\varphi$ are the nonzero components of potentials in
spherical coordinates. The magnetic field strengths are $F_{(1)ij}=F_{(3)ij}=
-\varepsilon_{ijk}\pa^k \f0 $, and in spherical coordinates have a single
 nonzero component $F_{\vartheta\varphi}= -P\sin\vartheta $.
  For the magnetic
extremal solution,  the components of the internal metric $G_{11}$ and $G_{22}$
are constants, and hence  perturbatively
\footnote{One-loop corrections are calculated with tree-level expressions.}
 $\D$ is also a
constant which we assume to be numerically small. In this case the effective
action (\ref{E10}) takes the form
\begin{eqnarray}
\label{E11}
S&=&\int d^4 x\sqrt{-g}\left(R-\frac{1}{2}(\pd\p)^2-(\pd\g)^2
\right. \nonumber \\
&-&
\left.\frac{1}{4}e^{2\g}(e^{-\p}+\D)F^2_{(1)}-\frac{1}{4}e^{-2\g}(e^{-\p}+\D)
F^2_{(3)}\right ).
\end{eqnarray}
\section{Solution of the equations of motion}
We look for a static spherically-symmetric solution of the field equations
of the loop-corrected action (\ref{E11}).
The general ansatz for the metric is
\begin{equation}
\label{E12}
ds_4^2 =- e^{\nu} dt^2 + e^{\l}dr^2 + e^{\mu}d\Omega_2^2.
\end{equation}
The field strength $F_{ij}$ is assumed to have the same functional form as
in the tree-level case: $F_{ij}= -\varepsilon_{ijk}\pa^k f(r)$, where the
function $f(r)$ is to be determined from the field equations. In spherical
coordinates this ansatz yields the only nonzero component
\begin{equation}
\label{E13}
F_{\vartheta\varphi}= f'(r)r^2\sin\vartheta.
\end{equation}
The equations of motion resulting from the action  (\ref{E11}) are
\begin{equation}
\label{E14}
D^2\phi + \frac{1}{4} e^{-\phi}(e^{2\g}F_{(1)}^2 +
e^{-2\g}F_{(3)}^2)=0,
\end{equation}
\begin{equation}
\label{E15}
\pd \left(\sqrt{-g}(e^{-\phi}
 +\D) e^{\pm 2\g} g^{\mu\mu'}g^{\nu\nu'}F_{(1,3)\mu'\nu'}\right) =0,
\end{equation}
\begin{eqnarray}
\label{E16}
R_{\mu\nu}-\frac{1}{2}g_{\mu\nu}R -\frac{1}{2} \left(\pd\p \pa_{\nu}\p
-\frac{1}{2}g_{\mu\nu}(\pa\p)^2 \right)
-\left(\pd\g \pa_{\nu}\g
-\frac{1}{2}g_{\mu\nu}(\pa\g)^2 \right) \nonumber \\
-\frac{1}{4}(e^{-\phi}
+\D)\left[ e^{2\g} \left(2(F_{(1)}^2)_{\mu\nu}-
\frac{1}{2}g_{\mu\nu}F_{(1)}^2 \right)
+ e^{-2\g} \left(2(F_{(3)}^2)_{\mu\nu}-\frac{1}{2}g_{\mu\nu}F_{(3)}^2
\right)\right]=0.
\end{eqnarray}
and
\begin{equation}
\label{Eg}
D^2\g -\frac{1}{4}(e^{-\phi} + \D )(e^{2\g}F_{(1)}^2 -
e^{-2\g}F_{(3)}^2)=0.
\end{equation}
At the tree level $\g = 0$ (see (\ref{E9})) and $F_{(1)}^2 =F_{(3)}^2$. In
the first order in $\D$, we have $\g  = \D \g_1 $.
Expanding all the expressions in
(\ref{Eg}) to the first order in $\D$, we obtain
$$
D^2 \g_1-\frac{1}{2} \left[e^{-\p}(F_{(1)}^2+F_{(3))}^2\right]_{(0)}\g_1=0.
$$
Here the subscript $(0)$ stands for the tree-level expressions.
From this equation it
follows that $\gamma_1=0$ is a solution.
Multiplying this equation by $\g_1$ and integrating over the space-time, we
obtain
$$
-\int d^4 x \sqrt g (D\g_1)^2=\int d^4 x\sqrt g \left(e^{-\p}F_{(1)}^2
\right)_{(0)} \g_1^2
$$
Since $F_{(1,3)}^2>0$ (see below), this equation can be satisfied only if
$\g_1=0$

Thus, in the first order in $\D$, we have $\g = 0$ also, and henceforth we
 set $\g =0$.

The field strengths  have the following nonzero components
$$
(F^2)_{\varphi\varphi} = (f'(r)r^2\sin\vartheta)^2 e^{-\mu}, \qquad
(F^2)_{\vartheta\vartheta} = (f'(r)r^2)^2 e^{-\mu},
$$
\begin{equation}
\label{E18}
F^2=2(f'(r)r^2)^2 e^{-2\mu}.
\end{equation}
Eq. (\ref{E14}) takes the form
\begin{equation}
\label{E19}
\p''-(\frac{\l'}{2}-\frac{\nu'}{2}-\mu')\p' + \frac{e^{\l-\p}}{2}F^2 =0.
\end{equation}
Away from the origin, Eq. (\ref{E15}) reduces to
\begin{equation}
\label{E20}
\pa_\vartheta\left(\sqrt{-g}(e^{-\phi}
+\D)g^{\vartheta\vartheta}g^{\varphi\varphi}F_{\varphi\vartheta}\right)=0.
\end{equation}
Since $\sqrt{-g} \sim \sin\vartheta,\quad g^{\varphi\varphi}\sim
{\sin}^{-2}\vartheta$ and $F_{\varphi\vartheta}\sim\sin\vartheta$,
Eq. (\ref{E20}) is satisfied identically.
To account for a singularity at the origin (point-like source of magnetic
charge), we calculate the flux of magnetic field through an arbitrary sphere
enclosing the origin
$$ \int_{S^2} F = \int d\vp d\vartheta\sin\vartheta F_{\varphi\vartheta} =
Const. $$
The Einstein equations (\ref{E16}) (with one index lifted) take the form
\cite{ll}:
\begin{equation}
\label{E21}
e^{-\l}\left(\mu'' + \frac{3}{4}{\mu'}^2 -\frac{\mu'\l'}{2}\right)-e^{-\mu}+
\frac{1}{4}e^{-\l}{\p'}^2 +\frac{1}{4}(e^{-\phi}+\D)F^2 =0,
\end{equation}
\begin{equation}
\label{E22}
e^{-\l}\left(\frac{{\mu'}^2}{2}
+\mu'\nu'\right)-2e^{-\mu}-\frac{1}{2}e^{-\l}{\p'}^2+
\frac{1}{2}(e^{-\phi}+\D)F^2=0,
\end{equation}
\begin{equation}
\label{E23}
e^{-\l}(2\mu''+2\nu''+{\mu'}^2+{\nu'}^2-\mu'\l'-\nu'\l'+\mu'\nu')
+e^{-\l}{\p'}^2 -(e^{-\phi}+\D)F^2=0.
\end{equation}

Our next aim is to solve the equations of motion to the first order in $\D$.
In the leading order ($\D=0$) we have
\begin{equation}
\label{E24}
{\nu}^{(0)} =-\ln\f0,\quad \l^{(0)} =\ln\f0,\quad \mu^{(0)} =\ln\f0 +2\ln r,
\quad
\p^{(0)} = \ln\f0,\quad {F^2}^{(0)} =2{q'}^2,\quad q' =\frac{\f0'}{\f0}.
\end{equation}
in the first order in  $\D$ we look for a solution in the form
\begin{equation}
\nu =-\ln\f0 +\D n ,\quad \l =\ln\f0 +\D l ,\quad \mu =\ln\f0 +2\ln r +\D m,
\label{E25}
\end{equation}
$$\p = \ln\f0 + \D\varphi,$$
$$F_{\varphi\vartheta} =P(1+\D p)\sin \vartheta.$$
Here $n,m,l$ and $\p$ are unknown functions, $p$ is a number. One also has
$$ \quad F^2 =2{q'}^2 (1+\D \tau),$$
where $\tau =2p-2m$.
In Appendix A we present the detailed solution of the equations of motion
assuming that $m = l=-n$ \footnote{There are four
equations (\ref{E19}) and (\ref{E21})-(\ref{E23}) for four unknown functions
$m,\, n, \, l$ and $\varphi$.  Our choice corresponds to the
requirement that in the first order in $\D$, as in the leading order, $\mu +
\nu = 2\ln r$.}.
 Although there are more equations than unknown functions, it
appears that the ansatz is consistent. We obtain the following solution:
\begin{equation}
\label{E26}
n=-m=-l = A_{-1}\frac{1}{\f0} + A_0 + A_1\f0
 +\frac{1}{2}\f0 \ln\f0
\end{equation}
and
\begin{equation}
\label{E27}
\vp =- A_{-1}\frac{1}{\f0} +A_0 +2p  +(A_1+ \frac{1}{2})\f0+
\frac{1}{2}\f0 \ln\f0.
\end{equation}
Here $A_i$ and $p$ are arbitrary constants.
The constants in the above solution are  partially constrained
 by imposing the
conditions that at large $r$ the metric is asymptotic  to the Minkowski
metric and that the asymptotic value of the dilaton is unity:
\begin{eqnarray}
A_{-1} + A_1 + A_0 =0,  \nonumber \\
 - A_{-1} +A_1 + A_0 +\frac{1}{2} +2p =0.
\label{E28}
\end{eqnarray}
From the relations (\ref{E28}) it follows that
\begin{equation}
\label{F1}
A_{-1} = p+\frac{1}{4}.
\end{equation}
\section{Horizon and matching of solutions at small and large distances}
Let us discuss our solution. The expressions (\ref{E26})-(\ref{E27}) were
obtained by making two expansions: (i) in $\D$, assuming that corrections
are smaller than the leading-order expressions: $|\ln f_0 |>|\D n |,\,|\ln
f_0 |>|\D l |$, etc., and (ii) by expanding the exponents $e^{\D n},\,e^{\D
l}$, etc. to the first order in $\D$ assuming that $1>|\D n |,\,|\D l |,
\ldots$. The bounds that define the range of validity of our
solution result both in  constraints on the free parameters $A_i$ and
on the domain of variation of $r$. \footnote{It is remarkable that solution
(\ref{E26})-(\ref{E27}) is obtained in an analytic form for all $r$ for which
we can use perturbation expansion in $\D$. Usually it is possible to solve
such problem only for small and large $r$ and to sew the asymptotic
solutions.}

Let us consider the $g_{00}$ component of the metric. Introducing
new variable $$x={\f0 (r)}^{-1} = \frac {r}{r+P},$$ we rewrite the
expression for the  metric component $g_{00}$ as
\begin{equation}
\label{D1}
e^{\nu}=  x\left[1+\D\left (A_{-1}(x-1)  +A_1(\frac{1}{x}-1) +\frac{1}{2x}
\ln\frac{1}{x}\right) \right].
\end{equation}
Here we used relations (\ref{E28}) to eliminate $A_0$.

The range of validity of the expression (\ref{D1}) is determined
by the inequalities
\begin{equation}
\label{D2}
1>\left |\D\left (A_{-1}(x-1)  +A_1(\frac{1}{x}-1) +\frac{1}{2x}
\ln\frac{1}{x}\right )  \right|
\end{equation}
and
\begin{equation}
\label{D3}
\ln \frac{1}{x}>\left |\D\left(A_{-1}(x-1)  +A_1(\frac{1}{x}-1)
+\frac{1}{2x}
\ln\frac{1}{x}\right)  \right|.
\end{equation}
Sufficient conditions for validity of the the inequalities
(\ref{D2}) and (\ref{D3})  are their validity for each separate term in
(\ref{D2}) and (\ref{D3}). Solving the inequalities, we obtain
\footnote{There is another
range $1 > A_1 > (e-1)^{-1}$ in which there exists a  solution
of the inequalities. However, for $A_1$ in this interval, the
range of small $x$  which we are interested in is excluded.}
\begin{eqnarray}
\label{D4}
\lefteqn{1>|\D A_{-1} |, \quad 1\gg |\D A_1 |}&&\nonumber \\
&&x>|\D A_1 |, \quad x> |\D|\ln\frac{1}{|\D|}.
\end{eqnarray}
Note that we assume that $|\D |\ll 1$.

Let us extrapolate the expression (\ref{D1}) to the region $x\sim
|\D |\ln\frac{1}{|\D |}$ and look for a zero (a would-be horizon) of the
function
$g_{00}$ at small $x$. For small $x$ we obtain the equation
\begin{equation}
\label{C1}
\frac{1}{\D } + A_{-1} =-\left ( \frac{A_1}{x} + \frac{1}{2x}
\ln\frac{1}{x} \right ).
\end{equation}
This equation has a solution (zero of the time component of the metric
$g_{00}$) provided $\D <0$.
Introducing new variable $y$ by the relation $x=e^{A_1 -y}$,
we transform Eq. (\ref{C1})
to the form $ye^y =A$, where $A=2\left(\frac{1}{|\D | } -
A_{-1}\right)e^{2A_1}$. For
$A\gg 1$ the asymptotic expansion of solution of this equation is
\cite{dbr}:
$$
y=\ln A - \ln  \ln A + \frac{\ln \ln A}{\ln A} + O\left( \frac{\ln \ln
A}{\ln A}\right)^2.
$$
An  approximate solution to  Eq. (\ref{C1}) is
\begin{equation}
\label{D5}
x_0 \approx \frac{A_1 + \ln\left(\frac{1}{|\D |} -  A_{-1}\right)}
{\frac{1}{|\D |} - A_{-1}}
= O\left(|\D | \ln\frac{1}{|\D |}\right).
\end{equation}

Let us consider the dilaton. In Eqs. (\ref{E14})-(\ref{E16}) the parameter $\D$
enters through the combination $\D + e^{-\p}$, where
\begin{equation}
\label{D6}
e^{-\p}=x\left [1-\D \left(A_{-1}(1-x) + (A_1 +\frac{1}{2}) (\frac{1}{x}-1) +
\frac{1}{2x}\ln\frac{1}{x}\right) \right].
\end{equation}
 Near the horizon, at $r_0\alt r$, the
function $e^{-\p}$ is of order $|\D | \ln\frac{1}{|\D |}$. In this
region, $\D + e^{-\p}$ and  $e^{-\p}$ are of the same order. At large $r$,
$e^{-\p}$ approaches unity and $e^{-\p} \gg \D$.
\footnote{It is important that we considered
only smooth decreasing functions $g_{00}$ and $e^\p$. The component of
the metric  $g_{rr} = e^\l = e^{-\nu}$ in the near-horizon region is a large
rapidly varying function.}
This suggests that we can
look for a solution of the field equations in the whole range
$r_0<r<\infty$, neglecting $\D$ as compared to $e^{-\p}$, which has a simpler
form than the loop-corrected solution and at the same time is sufficiently
close to the loop-corrected solution, in particular, has the horizon shifted
from the origin to $r>0$.

Let us consider a non-extremal solution of the
field equations (\ref{E14})- (\ref{E16}) at $\D =0$. The metric is
\cite{dlp}
\footnote{In ref. \cite{dlp}, as well as in related papers cited therein, the
action was obtained by dimensional reduction from higher dimensions. The 4D
vector fields and scalars appear as metric components with a 4D and internal
indices and with both internal indices respectively. In the case at hand, the
action (\ref{E11}) contains two scalar fields $\p$ and $\g$ and two vector
fields. Although one vector field comes from the metric and the other from
antisymmetric tensor, it can be verified explicitly that the action
(\ref{E11}) meets all the requirements of \cite{dlp}.}:
\begin{equation}
\label{D7}
ds^2 = -\left(1 +\frac{k\, {\sinh}^2 \mu}{r}\right)^{-1} \left(1-\frac{k}{r}
\right )dt^2 +
\left(1 +\frac{k\, {\sinh}^2 \mu}{r}\right)\left[\left(1-\frac{k}{r}
\right)^{-1} dr^2 +r^2d\Omega_2^2 \right],
\end{equation}
the dilaton and magnetic field strength are
\begin{equation}
\label{D8}
e^\p = \left(1 +\frac{k \,{\sinh}^2 \mu}{r}\right);\qquad
F_{(1)} =F_{(3)}=\frac{k\, \sinh{2\mu}}{2r^2}{\varepsilon}_2,
\end{equation}
where ${\varepsilon}_2$ is the volume of the unit 2-sphere, $k$ is the
position of  horizon.   The extremal limit
is $k \rightarrow 0$ and $\mu \rightarrow \infty$ with $ke^{2\mu}$ fixed.
 The $g_{00}$ component of the metric interpolates smoothly between
unity at $r=\infty$ and zero at the horizon. Next, we explore a
possibility to approximate the loop-corrected solution by a more simple
tree-level non-extremal magnetic black hole.

To find a non-extremal field configuration  which is sufficiently close
to the loop-corrected one, it is natural to
demand that (i) both solutions have equal magnetic charges, (ii) both
solutions have identical asymptotic behavior,
i.e. at large  $r$ the leading
$O(\frac{1}{r})$ terms of the metrics and dilatons of both solutions are
equal,
and (iii) locations of the horizons of both solutions are of the same order.
Let us discuss these requirements successively.

Equating the magnetic charges of both solutions, we have
\begin{equation}
\label{D9}
\frac{k}{2}\,\sinh \, 2\mu =P(1-|\D|p).
\end{equation}

The asymptotics of the component  $g_{00}$ of the loop-corrected metric
(other components differ from this expression by the sign of the
$O(\frac{1}{r})$ term) is
$$ g_{00}^{(1)} = 1- \frac{P}{r}[1+|\D|(A_1 -p +\frac{1}{4})] +
O(\frac{1}{r^2}),$$
where we used the relations (\ref{E28}) and (\ref{F1}).
The asymptotics of the corresponding component of the metric of non-extremal
solution is
$$ g_{00}^{(2)} = 1- \frac{P}{r}\left[1+ b +\frac{k}{P}\right]
+O(\frac{1}{r^2}),$$
where
$$b=\frac{k\,{\sinh}^2 \, \mu}{P} -1. $$
Assuming that the non-extremal solution is close to extremality, from
Eq.(\ref{D9}) we obtain
\begin{equation}
\label{D10}
b\simeq -\frac{k}{2P}-|\D|p.
\end{equation}
Equating the asymptotics of both metrics and substituting (\ref{D10}), we have
\begin{equation}
\label{D11}
A_1 =\frac{k}{2P|\D|} -\frac{1}{4}.
\end{equation}

Asymptotics of the functions $e^{-\p}$ are
$$
e^{-\phi^{(1)}}= 1- \frac{P}{r}[1-|\D|(A_1 +p +\frac{5}{4})] +
O(\frac{1}{r^2})$$
and
$$
e^{-\phi^{(2)}} = 1- \frac{P}{r}(1+ b)
+O(\frac{1}{r^2}).$$
Using (\ref{D10}), we have
$$
e^{-\phi^{(1)}}= 1- \frac{P}{r}[1-|\D|(\frac{k}{2P|\D|} +p +1)]$$
and
$$
e^{-\phi^{(2)}} = 1- \frac{P}{r}[1-|\D|(\frac{k}{2P|\D|} +p)].       $$
We see that although it is impossible to obtain strict asymptotic equality
$e^{-\phi^{(1)}}=e^{-\phi^{(2)}}$,  these expressions
are very close if $(\frac{k}{2P|\D|} +p) \gg 1$.

Further specification requires additional assumptions. Assuming that
locations of horizons of both solutions are of the same order, i.e. $k\sim
P|\D|\ln \frac{1}{|\D|}$, we have
\begin{equation}
\label{D12}
A_1 \sim \ln \frac{1}{|\D|},
\end{equation}
and
$$ {\sinh}^2 \, \mu \sim \frac{1}{|\D|}\ln \frac{1}{|\D|}. $$
For small $|\D|\ll 1$ we have ${\sinh}^2 \, \mu \gg 1$, i.e. the
non-extremal configuration is close to extremality.
 Taking the free parameter $p =O(1)$ and noting
 that $\ln \frac{1}{|\D|} \gg 1$, we find that
 both asymptotics of the metrics
and dilatons are very close to each other.

Assuming validity of the strict equality, $k =
r_0$, and taking $p\sim 1$,
from the equation for the horizon (\ref{D5}) we obtain
$$
k \simeq P\frac{A_1 + \ln \frac{1}{|\D|}}{\frac{1}{|\D|}},$$
where we neglected  the constants of order O(1) as
compared to $\frac{1}{|\D|}$. Substituting for $A_1$ the expression
(\ref{D11}),
 we have
$ k=2P|\D|\ln \frac{1}{|\D|}$.
\section{Thermodynamic properties}

Let us discuss some thermodynamic properties of the loop-corrected
configuration.  The  loop-corrected ADM mass  calculated from the asymptotics
of the the metric is
\begin{equation}
\label{E40} M=P\left[1+|\D|(-A_{-1}+A_1
+\frac{1}{2})\right ]
\end{equation}

To calculate the Hawking temperature we consider the near-horizon region where
we take $r= r_0 +{\rho}^2$ and, assuming that $ r_0 \gg {\rho}^2$, expand the
metric in powers of ${\rho}^2$. The  Hawking temperature is determined from
the requirement that the resulting metric has no conical singularity. We
obtain
\begin{equation}
\label{E41}
T_{bh}=\frac{1}{4\pi \left(1+\frac{|\D|}{r_0} \right)}.
\end{equation}
The geometric entropy defined as one fourth of the horizon area is
\begin{equation}
\label{E42}
S_{bh}= \pi x_0^2 \left (1+\frac{|\D|}{r_0} \right).
\end{equation}
Assuming the validity of the second law of thermodynamics, i.e. that  the
thermodynamic relation $\delta E_{bh}=T\delta S$
is approximately satisfied with $\delta S=S_{bh}$
\footnote{Note that at the string-tree level, at $\D =0$, the entropy of
magnetic black hole vanishes.}
 and $T=T_{bh}$, we obtain
$\delta E_{bh} \sim r_0 \sim P|\D|\ln \frac{1}{|\D|} $.
On the other hand, taking
\begin{equation}
\label{E43}
\delta E_{bh} \sim \delta M =P|\D|(-A_{-1}+A_1 +\frac{1}{2}),
\end{equation}
 we obtain that
$$ (-A_{-1}+A_1) \sim \ln \frac{1}{|\D|}$$
which is consistent with the estimate (\ref{D11}).

Let us discuss how the thermodynamic properties of the loop-corrected
solution  match those of the tree-level one.
The energy, charge and geometric entropy of the non-extremal magnetic
black hole  (\ref{D7})-(\ref{D8}) are \cite{dlp}
\begin{eqnarray}
E&\sim &k (3+\cosh 2\mu ), \nonumber \\
Q&\sim &k\sinh 2\mu,  \nonumber \\
S_{bh}&= &\pi k^2{\cosh}^2 \mu.
\label{E44}
\end{eqnarray}
Assuming that the non-extremal solution is close to extremality, i.e.
$\mu\gg 1$, the excess of energy above the extremal limit is
\begin{equation}
\label{E45}
\delta E \sim k.
\end{equation}
Supposing that the non-extremal solution is close to the loop-corrected
configuration in the sense discussed in the preceding section, and
substituting in the expression (\ref{E43}) for $\delta M$ the expressions
for $A_1$ and  $A_{-1}$ from (\ref{D10}), we obtain again that
$\delta M \sim k \sim P|\D|\ln \frac{1}{|\D|} $.
Finally, we can note that the Hawking temperature $T_{bh}=\frac{1}{4\pi k
\,{\cosh}^2 \mu} $ and the geometric entropy $S_{bh}=\pi k
\,{\cosh}^2 \mu $ of the non-extremal solution are close to the
corresponding loop-corrected expressions.

The entropy of an extremal dyonic black hole with nonzero electric
and magnetic charges is equal to $S_{bh} = \frac{\pi}{G_4} \sqrt{Q_1 Q_2 P_1
P_2} $.
In the non-extremal case the same formula for the geometric entropy is valid
with the charges $P_i = k\, \sinh {\gamma}_i \cosh {\gamma}_i$ and  $Q_i =
k\, \sinh {\d}_i \cosh {\d}_i$, $i=1,2$, where ${\gamma}_i$ and ${\d}_i$ are
boost parameters \cite{cvts2}. Extremal dyonic black hole is obtained in the
limit $k\rightarrow 0, \,
{\gamma}_i\rightarrow\infty,\,{\d}_i\rightarrow\infty$, with $P_i$ and $Q_i$
held fixed.
The solution (\ref{D7})-(\ref{D8})  can be
considered as a non-extremal counterpart of the extremal magnetic
solution of the chiral null model with two magnetic charges (\ref{E9}).
 The entropy of the non-extremal solution can be
obtained by the same substitution of charges as discussed above.
  Setting
in the expression for the entropy of the extremal dyonic black hole $Q_1
=Q_2 = k$ and $P_1 =P_2 =k\,{\cosh}^2 \mu \,$ we obtain for the entropy of
the
non-extremal magnetic black hole the following expression :
$$
S_{bh}=\pi k^2{\cosh}^2 \mu,
$$
which is the same as above (\ref{E44})

Statistical entropy of the non-extremal 4D dyonic black hole can be
calculated using the D-brane technique in the near-extremal limit where all
the charges are large \cite{lostmal}. In the case at hand, for non-extremal
magnetic black hole, we cannot repeat this calculation literally. If,
however, in the near-extremal limit, $k\rightarrow 0, \g \rightarrow\infty$
magnetic black hole can be substituted with dyonic black hole with the
electric charges $Q_1 =Q_2 =k$ and the D-brane counting could be extended to
this case, it would provide statistical origin of the entropy of the
non-extremal magnetic black hole. Anyway, the above remarks can be of
relevance in other approaches to counting of the statistical entropy
\cite{ss}.
\section{Discussion}
In this paper we have discussed a solution to the equations of motion of the
string-loop-corrected effective action in the simplest setting: when the
moduli fields and therefore the string-loop correction to the gauge
couplings are
constant. As the tree-level solution we chose the magnetic black hole given
by the chiral null model.

The main result of our study is that  the
horizon of the loop-corrected solution is moved away from
the origin and that
the loop-corrected solution can be approximated by  a non-extremal black
hole.
It was found that the requirements that both the loop-corrected and the
non-extremal solutions have the same large-$r$ asymptotics and positions of
their horizons are of the same order can be fulfilled if the free parameters
of both solutions are connected in a special way.

 The string-tree-level chiral null model provides a  solution to
 the low energy effective action  which in a special
renormalization  scheme  receives no
$\a'$ corrections \cite{bh2}. The solution of the one-loop-corrected
effective action we considered
(\ref{E10}) is no longer expressed in terms of  harmonic functions and the
$\a'$-corrections are present.
However, because now the horizon is shifted away from the origin, it is
possible that the $\a'$ corrections are small and can be treated
perturbatively.
\acknowledgments
This work was  partially supported by the RFFR  grant No 98-02-16769.
\appendix
\section{Solution of field equations}
In the first order in parameter $\D$ the Einstein equations
(\ref{E21})-(\ref{E23}) are
\begin{equation}
\label{A1}
m''+m'(q'+\frac{3}{r})-l'(\frac{1}{2}q'+\frac{1}{r})
+{\vp}'\frac{q'}{2} -\frac{l-m}{r^2}+\frac{1}{2}{q'}^2 s =
-\frac{1}{2} \f0 {q'}^2,
\end{equation}
\begin{equation}
\label{A2}
m'\frac{2}{r}+n'(q'+\frac{2}{r})-{\vp}'q'-2\frac{l-m}{r^2}+{q'}^2 s  =
 -\f0 {q'}^2,
\end{equation}
\begin{equation}
\label{A3}
m''+n'' +m'\frac{2}{r} -l'\frac{1}{r} + n'(-q' +\frac{1}{r}) +
{\vp}'q' -{q'}^2 s = \f0 {q'}^2.
\end{equation}
Here
\begin{equation}
\label{A4}
s=l-\vp +\tau =l-\vp -2m +2p
\end{equation}
We also need the equation for the dilaton (\ref{E14}) in
the  $O(\D)$ order:
\begin{equation}
\label{A5}
{\vp}''+\frac{2}{r}{\vp}'+\frac{1}{2}(2m'+n'-l')q' +{q'}^2 s =0.
\end{equation}
We look for a solution such that $m=l=-n$, because in this case, as at the tree
level, the components of the metric satisfy the relation $g_{tt} =
 g^{-1}_{rr}$. Substituting this ansatz in equations (\ref{A1})-(\ref{A5}),
we have
\begin{equation}
\label{A6}
m''+m'(\frac{1}{2}q'+\frac{2}{r})
 +{\vp}'\frac{q'}{2} +\frac{1}{2}{q'}^2 s =
-\frac{1}{2} \f0 {q'}^2,
\end{equation}
\begin{equation}
\label{A7}
m'q'+ {\vp}'q' - {q'}^2 s =  \f0 {q'}^2,
\end{equation}
and
\begin{equation}
\label{A8}
{\vp}'' +\frac{2}{r}{\vp}'+{q'}^2 s = 0,
\end{equation}
where
$$s=2p-m-\vp
$$
Eqs. (\ref{A2}) and  (\ref{A3}) reduce to the same Eq.  (\ref{A7}).
Combining  the equations as (\ref{A6})-(\ref{A7})+(\ref{A2}),
substituting the expression for $s$ , and introducing new variable
$u=m+\vp$, we obtain
\begin{equation}
\label{A9}
u''+u'(q'+\frac{2}{r})-{q'}^2(u -2p)=0
\end{equation}
\begin{equation}
\label{A10}
u'+q'(u -2p)=\f0 {q'}
\end{equation}
and
\begin{equation}
\label{A11}
{\vp}'' +\frac{2}{r}{\vp}' -{q'}^2 (u -2p)=0
\end{equation}
Eqs.(\ref{A9}) and (\ref{A10}) are not independent: the first one can be
obtained from the second one as
$$(\ref{A10})' + \frac{2}{r}(\ref{A10}) = (\ref{A9})
$$
 Thus, finally we  obtain the following system
\begin{equation}
\label{A12}
u'+q'u = (\f0 +2p)q'
\end{equation}
and
\begin{equation}
\label{A13}
{\vp}'' +\frac{2}{r}{\vp}' +(2p-u){q'}^2 =0
\end{equation}
Eq. (\ref{A12}) can be  integrated yielding
$$ u= \frac{C}{\f0}+\frac{\f0}{2}+2p, $$
where $C$ is an arbitrary constant.
Using this expression we integrate (\ref{A13}):
\begin{equation}
\label{A14}
\vp = C_2+C_1 \f0 + \frac{C}{2\f0} +\frac{1}{2}\f0 \ln \f0
\end{equation}
Finally, for $m=-\vp +u$ we obtain
\begin{equation}
\label{A15}
m =- C_2+2p +(C_1-\frac{1}{2}) \f0 + \frac{C}{2\f0} -\frac{1}{2}\f0 \ln \f0.
\end{equation}
The relation between the constants used here and those in the main test is:
$A_0=C_2-2p$; \, $A_1=C_1-1/2$; \, and $A_{-1}=-C/2$.

\vskip2.mm

\end{document}